\documentstyle[12pt]{article}
\begin{document}
%
%
%
%
\title
{Multifractal Structure of the Sea-Bottom Topography
in the Korean Sea}
\author
{ Kyungsik Kim\thanks{E-mail: kskim@dolphin.pknu.ac.kr}  
and Y.S. Kong$^1$ \\
$ Department$ $of$ $Physics$, $Pukyong$ $National$ $University,$ \\
$ Pusan$ $608-737$, $Korea$\\ 
%
$^1$ $School$ $of$ $Ocean$ $Engineering$, $Pukyong$ $National$ $University$,\\
$Pusan$ $608-737$, $Korea$}
\date {}
\maketitle\abstract
{The scaling behavior of the mutifractality for the sea-bottom topography
of the South Sea in Korea is numerically investigated.
In particular, we focus on the behavior of the $q$th-moment depth-depth correlation function 
of the sea-bottom topography and its multifractal spectrum.
Through the multifractal analysis, the fractal dimension and the scaling exponents are  
obtained numerically, and the relation 
between the Hurst exponent and the fractal dimension is also derived. 
}
\vspace{0.5cm} 
Keywards : sea-bottom  topography, depth - depth correlation function \\
\hspace*{2.1cm}multifractals, generalized dimension, spectrum \\ 
\newpage
\indent
For more than two decades, there has been considerable interest in
the investigation of the scaling behavior on fractal models.
The pioneering work$[1]$ on fractals has introduced the concept of fractals
and shown some relation between the self-similar fractals and the self-affine
fractals.   
The self-affine fractals$[2-5]$ constitute 
random and complicated structure that has been applied to broader range of problems
such as Eden and ballistic deposition model$[6-9]$, mountain heights, 
clouds, coast lines and cracks.
Among other examples of many fractal models we also can mention 
the self-avoiding random walk, percolation clusters, diffusion-limited aggregations, 
random resistor networks, polymer bonds, turbulences, chaotic motions
$[2,4,5,10-13]$, etc.  \\
%
%
\indent     
On the other hand, Matsushita and Ouchi$[14]$ have recently shown 
that the self-affine of  mountain topography in Japan is mainly obtained by using
the numerical method, from which they have discussed the relation $D_f=2-H=1.37$ between the 
self-affine exponent $H$ and the fractal dimension $D_f$ for the standard deviation of 
mountain heights. More recently, Katsuragi and Honjo$[15]$ have dealt with self-affine
fractal profiles on the entropy spectra method and for the  whole range of a mountain they 
have obtained the generalized multi-affine profiles and the entropy spectra. \\
\indent  
It has recently discussed that  
the multifractals$[2,5]$ have essentially the distributions of singularities of 
the scaling exponents
on interwoven sets of various fractal dimensions. 
The multifractal formalism for the box-counting method is represented 
in terms of their generalized dimensions and spectra.  
The multifractality of self-affine fractals has been discussed by Barab$\acute{a}$si 
$et$ $al$$[16,17]$.
They have investigated the multi-affine function and the multifractal spectra, and calculated      
the generalized Hurst exponent from the $q$-th  order height-height correlation function.     
Very recently, the generalized dimension and the scaling exponent 
which are interesting in multifractals for mountain heights in Korea have been calculated
by Kim and Kong$[18]$.
We think that it is interesting to be extended to the multifractals,
although we do not have enough sets of data for the seabottom depth.  
In particular, the multifractal theory is a good basis for analyzing the statistical
data on the seabottom depth, and really provides the probabilistic evidence
for the relation between Hurst exponent and the fractal dimension.\\
\indent
The main purpose of this paper is to investigate numerically the behavior
of the depth-depth correlation function and its multifractal spectrum in the 
seabottom depth from the seabottom topography
between the longitude $129^{o} 21^{\prime}$ - 
$129^{o} 54^{\prime}E$ and the latitude $34^{o} 52^{\prime}$ -  $35^{o} 19^{\prime} N$
located near to the east of Pusan. 
The primary reason we select this region is
that this region has more data
for sea-bottom depth than those of any other region around Korean peninsula. 
We consider well known relations of 
the depth-depth correlation function and the multifractals. 
The $q$th-moment depth-depth correlation function is also numerically
discussed from the multi-scaling properties of seabottom depth function. 
In addition, from the extrapolated sets of data for the seabottom depths, 
we briefly 
explain an efficient and convenient method
relevant to numerical investigation on the scaling behavior of the generalized dimension 
and the spectra. \\
%
%
%
%
\indent 
First of all, we are interested in the scaling quantities 
obtained from the extrapolated data of the seabottom depth   
on a two dimensional square lattice between 
$129^{o} 21^{\prime}$ - $129^{o} 54^{\prime} E$ and $34^{o} 52^{\prime} $ - $35^{o} 19^{\prime} N$,
as shown in Fig. $1$.
Moreover, the current data are taken for the values of the seabottom depths projected on 
$100\times100$ lattice points in square area of $50\times50$ $km^2$,
where the distance between one lattice point and its neighboring lattice point
is $500$ $m$. 
Let's consider the $q$th-moment depth-depth correlation function $C_q (r)$
defined by
\\
\begin{equation}
C_q  (r) = \frac{1}{N}{\sum_{i=1}^{N}} \vert D( r_i + r) - D( r_i )\vert^q ,
\label{eq:a1}
\end{equation}                                        
\\
where $D( r_i)$ is the $i$th depth in the $r$-direction and   
$r$ denotes two perpendicular longitudinal or latitudinal direction 
on a two dimensional square lattice.
This correlation function can be described  as a non-trivial multi-scaling behavior;
\\
\begin{equation}
C_q(r) \sim r^{qH_q} , 
\label{eq:b2}
\end{equation}                                        
\\    
where $H_q$ is the generalized Hurst exponent in the limit of $r\rightarrow 0$.\\ 
\indent 
Next, we discuss the generalized dimension $D_q$, the scaling exponents
$\alpha_q$, and $f_q$ on the multifractal structure.
In general, it has been known that the generalized dimension is represented as 
the fractal distribution having the singular values infinitely. 
Accordingly, the generalized dimension and the scaling exponents are known
to be given by following equations$[4,5]$
\\
\begin{equation}
D_q ={ \lim_{\epsilon\rightarrow0}^{}}\frac{1}{q-1}\frac{\ln {\sum_{i}^{}} n_i {p_i}^q}
{\ln\epsilon} ,
\label{eq:e5}
\end{equation}                                        
\\    
\begin{equation}
\alpha_q ={ \lim_{\epsilon\rightarrow0}^{}}\frac{1}{\ln\epsilon}\frac{{\sum_{i}^{}} 
{p_i}^q ln p_i}{{\sum_{i}^{}} n_i {p_i}^q},
\label{eq:f6}
\end{equation}                                        
\\
and                                                            
\begin{equation}
f_q = \lim_{\epsilon\rightarrow0} \frac{1}{\ln\epsilon}[\frac{\sum_{i} 
{p_i}^q {\ln} {p_i}^q}{\sum_{i} n_i {p_i}^q} - \ln \sum_{i} n_i {p_i}^q] ,
\label{eq:g7}
\end{equation}                                        
\\    
where $p_i$ is the probability of the seabottom depth existing on the $i$-th box with 
the square area of  $\epsilon\times\epsilon$ and the scaling quantity $n_i$ is the number of the box 
having the probability $p_i$. By introducing the above expressions, the spectra
$f_q$ and $\alpha_q$ are simply calculated from eqs.$(4)$ and $(5)$
with $f_q =q\alpha_q -(q-1)D_q$ and $\alpha_q = \frac{d}{dq}[(q-1)D_q] $,
where $f_q$ and $\alpha_q$ can be obtained by Legendre transformation. 
In our scheme, we will make use of eqs.$(3)$-$(5)$ to find out
the fractal dimension and other scaling exponents,
and these mathematical techniques lead us to
more general results.
\\
%
\indent
We present more detailed numerical data of the depth-depth correlation function
of the seabottom depth and its multifractal spectrum.
As mentioned above,
we assume that the seabottom depths divided by the intervals of $500$ $m$ are located 
on each region of two  dimensional lattice. As listed in Table $1$, 
we take into account the data of three regions $A$ - $C$ in seabottom depths
between $110$ $m$  and $125$ $m$.
As we only restrict ourselves to the longitudinal direction from our data, 
the generalized Hurst exponent from the $q$th-moment depth-depth
correlation function is calculated numerically. Hence, from eq.$(2)$
the generalized Hurst exponents 
$H_2$, $H_3$, and $H_4$ on the longitudinal direction
are found to be $0.599$, $0.587$, and $0.586$, respectively. \\
\indent
In our multifractal structure,
the box-counting method is used for two square areas
of $2.5\times2.5$ $km^{2}$ and $5\times5$ $km^{2}$.
Through the multifractal analysis, the generalized dimension and
the scaling exponents are also obtained from the data with the number of seabottom depths 
in Table $1$. 
These values by using the theoretical
expressions of eqs.$(3)-(5)$ are calculated numerically
in our three regions. In Fig. $2$ we show 
the scaling exponents
$\alpha_q$ and $f_q$ on the multifractal structure,
and it is found numerically
that the scaling exponents are estimated as $\alpha_{+\infty} = 1.097696$ and   
$\alpha_{-\infty} = 2.172183$ only in the case of the region $A$.
Particularly, as shown schematically
in Fig. $3$ and Table $2$, it is obtained that the maximum values of the generalized dimension,
i.e., in our three regions $A$, $B$, and $C$ the fractal dimension $D_f = D_0
$ are respectively calculated as $1.312476$,
$1.366726$, and $1.372243$
for the case of the square area $2.5\times2.5km^2$. 
Hence, we obtain that Hurst exponent takes approximately $0.65$
in our multifractal structure,
while it takes the value $H$=$0.63$ in Japanese mountain topography$[14]$.\\
\indent
     In conclusion, we have remarkably studied on the $q$th-moment depth-depth 
correlation function and
its multifractal spectrum in the 
seabottom topography of the South Sea in Korea, as shown in Fig.$1.$
Specifically,
the multifractals have been investigated numerically 
by employing the box-counting method on a two dimensional square lattice. 
Based on these results, we expect that  further analytical and numerical progress for
multifractals may be achieved from more data for the ocean floor
on diverse region of seabottom topography.           
Futhermore, it will be useful to apply and 
to reinvestigate the above formalism in many scientific fields$[19,20]$ such as  
quantum disorder systems, irreversible growth structures,
and fractured surfaces of minerals. \\
\indent 
This work was supported in part by the academic research fund
of Pukyong National University in Korea. \\
%
%
%

%
%
%
%
\newpage
\section*{Figure Captions}                                            
\vspace{0.5cm}
Fig. $1$ $:$  The seabottom depth in seabottom topography taken from $\frac{1}{312000}$ scale map 
between the longitude $129^{o} 21^{\prime}$ - $129^{o} 54^{\prime}E$
and the latitude $34^{o} 52^{\prime}$ - $35^{o} 19^{\prime} N$, 
where the interval of seabottom depth between the solid lines is found to be $10m$.\\
\\
Fig. $2$ $:$ $f_q$ versus $\alpha_q$ by using the box - counting method for the case of 
the square area $2.5\times2.5km^{2}$. The
values of the seabottom depths in three regions $A, B,$ and $C$ are respectively
given by the thin solid, dashed, and thick solid lines. \\ 
\\
Fig. $3$ $:$ Plots of $\alpha_q$ and $f_q$
as a function of $q$ on the seabottom depths 
for the case of the square area 
$2.5\times2.5 km^{2}$,
where the vertical bars are error bars averaged over the square areas
of three regions $A, B,$ and $C$.\\
\\
\section*{Table Captions}
\vspace{0.5cm}
Table $1$ $:$  Number of the seabottom depths in our three regions $A$ - $C$ 
between $110m$ and $125m$ on the seabottom topography.  \\ 
\\
Table $2$ $:$ Summary of values of $D_q$, $\alpha_q$, and $f_q$ calculated from the data of three 
regions $A$, $B$, $C$ for the case of the square area $2.5\times2.5km^{2}$.
\end{document}